\documentclass[useAMS,usenatbib,usegraphicx]{mn2e}
\title[Rotation of the cluster of galaxies A2107]{Rotation of the cluster of 
galaxies A2107}
\author[M. Kalinkov et al]
{M. Kalinkov$^{1}$\thanks{E-mail:
markal@astro.bas.bg}, T. Valchanov$^{1}$\thanks{E-mail:
tony@astro.bas.bg}, I. Valtchanov$^{2}$\thanks{E-mail:
i.valtchanov@imperial.ac.uk}, I. Kuneva$^{1}$
M. Dissanska$^{1}$
\\
$^{1}$Institute of Astronomy, 72 Tsarigradsko Chauss\'{e}e blvd, Sofia 1784, 
Bulgaria\\
$^{2}$Blackett Laboratory, Imperial College, Prince Consort Road, 
London SW7 2BW, United Kingdom}

\begin{document}

\date{Accepted \today. Received \today; in original form 
\today}


\maketitle

\label{firstpage}

\begin{abstract}

We present indications of rotation in the galaxy cluster A2107 by a
method that searches for the maximum gradient in the velocity field in
a flat disk-like model of a cluster. Galaxies from cumulative
sub-samples containing more and more distant members from the cluster
centre, are projected onto an axis passing through the centre and we
apply a linear regression model on the projected distances $x$ and the
line-of-sight velocities $V$. The axis with the maximum linear
correlation coefficient $r_{max} = \max \left[ r(V,x) \right]$ defines
the direction of the maximum velocity gradient, and consequently it
presents the major axis of the apparently elliptical cluster.  Because
the effects of rotation are subtle, we put strong emphasis on the
estimation of the uncertainties of the results by implementing
different bootstrap techniques. We have found the rotational effects
are more strongly expressed from distances $0.26 \div 0.54$ Mpc from
the cluster centre. 
The total virial mass of the cluster is
$(3.2\pm0.6)\times10^{14} {\cal M}_{\sun}$, while the virial mass,
corrected for the rotation, is $(2.8\pm0.5) \times 10^{14}{\cal
  M}_{\sun}$.

\end{abstract}


\section{Introduction}

Velocity gradients suggestive of galaxy cluster rotation were found in 
several studies
(e.g. Kalinkov 1968, Gregory 1975, Gregory \& Tifft
1976, Gregory \& Thompson 1977, Materne \& Hopp 1983, Materne 1984,
Williams 1986, Oegerle \& Hill 1992, Sodr\'{e} et al 1992,  
Biviano et al. 1996, Tovmassian 2002). 
It is commonly accepted opinion that there is no evidence of rotation of the
galaxy clusters, however, if the galaxy clusters rotate their dynamics
and dynamical evolution would be different. The velocity dispersion
profiles must be corrected for the rotation and the corresponding
virial mass estimation will be different. Many theoretical
constructions are based on the assumption of no rotation -- e.g. the
infall models and especially the theory of caustics (Reg\"{o}s \&
Geller 1989, Diaferio \& Geller 1997, van Haarlem et al. 1993, Diaferio
1999, Rines et al. 2003).

Here we make an another attempt to reveal rotation in the galaxy
cluster A2107. We have chosen this particular cluster to illustrate
our method, because it is well studied and indications of rotations
were already found (Oegerle \& Hill 1992).

The structure of the paper is as follows: in Section 2 we present our
method, the different bootstrap techniques that we use to estimate the
uncertainties of the derived rotational parameters are presented in
Section 3.  In the next Section 4 we present the data for A2107 and in
Section 5 our results. We finish with discussion and conclusions
(Section 6).  
In order to compare our results with the previous
studies we assume an Einstein-de Sitter cosmology and $H_{0} = 100$ km
s$^{-1}$ Mpc$^{-1}$. For this model at the cluster redshift $z=0.04100 << 1$, 
$10\arcmin$ correspond to 358 kpc.

\section{Method}
\label{sec:method}

We consider a flat, disk-like galaxy cluster with regions with nearly
solid body rotation. The main idea is to find the axis of the maximum
velocity gradient which, in the disk-like model, defines the major
axis of the cluster.  The minor axis is the axis of rotation.

It is known the rotational effects are weak and their search would be
successful if some cumulative technique is applied. Let us take
sub-samples $\{i\}$ of the cluster member galaxies, arranged by their
projected distance from the cluster centre. Thus any sub-sample
$\{i\}$ contains the first $i$ galaxies out to a given projected
distance $d_{i}$. The sub-samples $\{i\}$ are not independent.

Let us introduce an axis passing through the cluster centre and
rotating around it and we define the positional angle $\varphi$ in the
usual way (anticlockwise from N) as shown on Fig.~1. 
We assume the axis has direction in sense SN to WE, NS, EW. In this
case the positional angle is $0 \degr \leq \varphi < 360 \degr$.

\begin{figure}
\includegraphics[width=8cm]{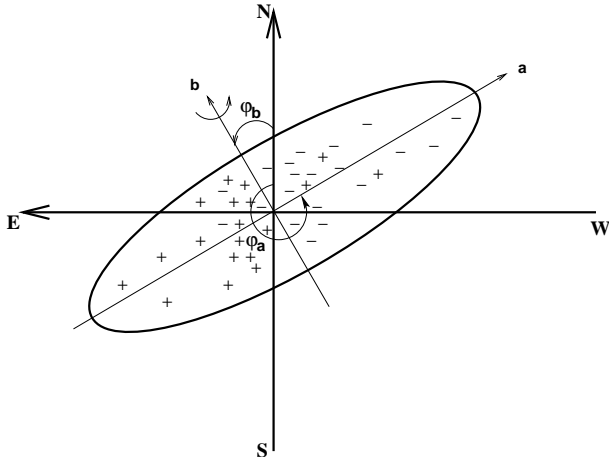}
 \caption{A schematic and idealistic representation of a disk-like
   cluster that indicates the axes and the angles used in the
   method. The symbols ``+'' and ``-'' are velocity deviations of the
   galaxies from the mean cluster velocity. 
   Note the positional angle of rotation
   $\varphi_b$ spans the whole range from $0$ to $360$ degrees. }
 \label{fig:scheme}
\end{figure}

We project the galaxies from sub-sample $\{i\}$ onto an axis with
positional angle $\varphi$ and examine the linear regression
\begin{equation}
 V = \alpha + \beta(x - \langle x\rangle),       
 \label{eq:fit}
 \end{equation}
where $V$ is the observed line-of-sight velocity, $x$ are the
projected cluster-centric distances on the chosen axis,
$\langle x \rangle$
is the mean of $x_{i}$ for the sub-sample $\{i\}$, and in fact 
$\alpha \equiv \langle V \rangle$.

Our aim is to find the maximum gradient in the velocity field. In
linear approximation, this corresponds to the axis with
$\varphi_{rmax}$ for which the linear correlation coefficient between
$V$ and $x$ is maximal, $r_{max}$. Consequently, the standard
deviation (st.dev.) of the regression for the sub-sample \{i\}
\begin{equation}
s = \left(\sum_{k=1}^{i} (V_k - \langle\hat{V_i}\rangle)^{2}/(i - 2)\right)^{1/2}
\end{equation}
is minimal and we denote it by $s_{rmax}$ which corresponds to
$r_{max}$ for the same $\varphi$; $\hat{V_k}$ is the line-of-sight
velocity estimate, derived from the regression.
 
If the velocity field of a flat cluster is influenced by rotation then
the maximum velocity gradient will be along the major axis $a$ (or
equally, along the axis of maximum elongation).  The positional angle
of the maximum velocity gradient is $\varphi_{rmax} = \varphi_{a} =
\varphi_{smin}$, where the indexes $rmax$ and $smin$ denote the
maximum correlation coefficient and the minimum standard deviation of
the regression estimate. Correspondingly, the axis of rotation is at
$\varphi_{r0} = \varphi_{b} = \varphi_{smax}$, where $r0$ denotes zero
correlation coefficient. Note that there are two positional angles
where $r$ is zero. We define the position of the minor axis with
$\varphi_{r0} > \varphi_{rmax}$.

Searching for the major axis or maximum elongation axis with a limited
number of cluster members is prone to large uncertainties and few
interlopers could bias the estimated ellipticity and shape. Much
better approach would be by using X-ray observations because the hot gas
maps better the cluster potential and subsequently the cluster shape. 
Unfortunately the positional angles of the maximum elongation
in the cluster cores are known to vary depending on the X-ray isophote
level.

The virial mass ${\cal M}$ of a cluster is defined through the
velocity dispersion $\sigma_{V}$. In fact,
\begin{equation}
\sigma_{V} = s_{r0}((i-2)/(i-1))^{1/2}
\end{equation}
If the cluster rotates the velocity
dispersion has to be corrected ($\sigma_{V}^{c}$), according to the
formula above, but with $s_{rmax}$ instead of $s_{r0}$. Then ${\cal
  M} - {\cal M}^{c} = \Delta {\cal M}$ is a fictitious mass, due to
rotation.
 
Supposing that some noticeable effects of rotation really exist, then,
with our approach, we expect the following indications:
\begin{enumerate}
\item The correlation coefficient $r_{max}$ will be significant for at
  least few consecutive sub-samples $\{i\}$.
\item The positional angles $\varphi_{rmax}$ for these consecutive
  sub-samples will not be randomly distributed along the entire range
  of $\varphi$, but in a relatively narrow interval.
\item The variation of $r$ and $s$ for $0\degr \leq \varphi <
  360\degr$ would be close to sine waves and, for sub-samples $\{i\}$
  with significant $r_{max}$, the phase shifts of both curves should
  be according to (ii), in a narrow range.  The expressions for the
  sine waves of $r$ and $s$ are
 \begin{equation}  
  r(\varphi) = r_{max} \sin(\varphi -
  \varphi_{rmax}) 
 \end{equation}
 and
 \begin{equation}    
  s(\varphi) = (s_{r0} + s_{rmax})/2 + [(s_{r0}
    - s_{rmax})/2] \sin(\varphi - \varphi_{r0})
 \end{equation}   
\item The virial mass for some sub-samples $\{i\}$ will be
  significantly different as computed for $s_{r0}$ and $s_{rmax}$.
\end{enumerate} 

\section{Uncertainties, bootstrap techniques and randomisation.}
 \label{sec:boot}
 
The question of uncertainties is a crucial one because the effects of
rotation are weak. Where it is possible we rely on bootstrap
uncertainties.

We apply two bootstrap techniques. The first one is the standard or
the classical bootstrap (B) - (Efron 1982, Efron \& Tibshirani
1986). Shortly, for each sub-sample we find the positional angle
$\varphi_{rmax}$ and then resample the velocities $V$ and the
projected distances $x$ on this axis in order to derive the bootstrap
uncertainties of $\alpha,\ \beta,\ r,\ s$. This procedure however has
the disadvantage that it is impossible to find the uncertainties of
$\varphi_{rmax}$ and $\varphi_{r0}$.

That is why we apply simultaneously a modified bootstrap (MB)
techniques: the positions and velocities of sub-sample $\{i\}$ are
subject to bootstrap resampling after which the search for $r_{max}$
is carried out and the corresponding
$\varphi_{rmax},\ \alpha,\ \beta,\ r,\ s$,... are derived. Note that
each bootstrap sample leads to completely different
$r_{max},\ \varphi_{rmax},\ \alpha,\ \beta$,...

In all cases, we run $10^{4}$ bootstrap generations and calculate the
corresponding quantities for $\varphi [0\degr - 360\degr)$ with a
  decrement of $0 \fdg 1.$ We use the resulting distribution of the
  quantity in question to derive the confidence interval.

Let us denote with $w_{B}$ whatever of the quantities we discuss.  The
first estimator of the uncertainty we use is the bias-corrected (bc)
$68 \%$ confidence interval $w \in [w_{bc} (0.16), w_{bc} (0.84)]$
(Efron \& Tibshirani 1986) with
\begin{equation}
w_{bc}(t) = G^{-1}\left\{\Phi \left[\Phi^{-1}(t) +
  2\Phi^{-1}(G(\langle w_{obs} \rangle)) \right]\right\},  
\end{equation}
where $G$ is the cumulative distribution function (CDF) of the $w_B$,
$\langle w_{obs} \rangle$ is the observed mean value, and $\Phi$
is the CDF of the normal distribution (thus $\Phi^{-1}(t) = -1,+1$ for 
$t=0.16$ and $t=0.84$ respectively).
This uncertainty is $\Delta w_{ET}$. Efron \& Tibshirani (1986) uncertainty is 
effectively used by Shepherd et al. (1997) and Kalinkov, Valtchanov \& Kuneva
(1998a) for the space correlation functions of galaxies and clusters of 
galaxies. 

The second estimator of the uncertainty we implement is the bootstrap
standard deviation (Ling, Barrow \& Frenk 1986)
\begin{equation}
\Delta w_{B} = \left[ \sum_{m=1}^{10^{4}}(w_{B,m} - \langle w_{B}
  \rangle)^{2} /10^{4}\right]^{1/2},   
\end{equation}
where $\langle w_{B}\rangle$ is the mean bootstrap value.
This st.dev. is very effectively used by Kalinkov \& Kuneva (1986).

We have found that both uncertainties $\Delta w_{B}$ and $\Delta
w_{ET}$ are consistent with each other, but we prefer $\Delta w_{ET}$,
since the estimator is valid even if the distribution of $w$ is not
Gaussian. 

The uncertainty on the correlation coefficient can also be
computed using the classical Fisher z-transformation (e.g. Press et
al. 1992, Section 14.5). The uncertainties derived by this method are
also consistent with the other two.

There is one problem -- the bootstrap technique does not work when we
determine the uncertainties of the virial mass estimates.  In the
resampling, some objects are duplicated and consequently the distance
between them is zero and the cluster potential tends to
infinity.  In this case only we apply the jackknife estimate (see
e.g. Efron 1982).

In order to assess the significance of the derived quantities we use
randomised sub-samples for which the azimuthal angles (or $\varphi$)
of the galaxies centres are made randomly distributed in the interval
$[0\degr, 360\degr)$.  This new sample is subject to the same
  procedure of parameter and confidence interval estimates. Here the
  uncertainties are only $\Delta w_{B}$.  It is worth to note that
  with this procedure the observed radial density and velocity
  distributions are not altered.

\section{Data}
\label{sec:data}

Coordinates and heliocentric velocities of galaxies in the direction
of A2107 are taken predominantly from the most complete optical study
of Oegerle \& Hill (1992) -- further on OH.  Some corrections and
additions are made according to NED (new redshifts and coordinates)
and therefore new observational errors are defined. 
We use also data from other sources, e.g. Zabludoff et al.(1993). 
We reckon that galaxy Nr. 252 is the same as Nr. 273 (LEDA 94259) from
the list of Oegerle \& Hill (1992).
     
We adopt the X-ray centre of A2107 according to Ebeling et al. (1996)
RA $= 15^{h}39^{m}38 \fs 0$ and Dec $= 21\degr 47\arcmin 20\arcsec$
for J2000. Our results however do not substantially differ neither if
we adopt the original cluster centre of Abell (1958) and Abell, Corwin
\& Olowin (1989) nor if we take the cD galaxy (UGC 9958) as centre.

The member galaxies were selected using  ROSTAT
(Beers, Flynn \& Gerhardt 1990) as well as the cone diagrams and
the velocity distribution. There are 70 galaxies out to 37.0 arcmin
from the X-ray centre within $\langle V \rangle \pm 2 \sigma_{v}$.
  Zwicky galaxy 136-22 is the farthest member.
For the non relativistic case (i.e. in the observed rest frame), the 
mean velocity is $\langle V\rangle =
12291^{+73}_{-75}$ km s$^{-1}$, the standard deviation
$627^{+56}_{-41}$ km s$^{-1}$, the bi-weighted estimate of location
and scale $V_{bw} = 12318$ km s$^{-1}$ and $s_{b} = 639$ km s$^{-1}$,
the median $V_{m} = 12380$ km s$^{-1}$ and the normalised absolute
deviation $s_{m} = 623$ km s$^{-1}$, skewness -0.28 and kurtosis
-0.40.
 
 All our results cited here are for this sample with size $n = 70$.

There are two galaxies at 1.2 arcmin from the cluster centre (Nrs. 242
and 289 from the list of OH) with velocities 14069 and 14028 km
s$^{-1}$. These galaxies could be taken as cluster members if we stand
on the infall theory, since the velocity dispersion at the cluster
centre should be the largest.  Including these two galaxies we obtain
for $n = 72$: $\langle V\rangle = 12 339^{+79}_{-80}$ km s$^{-1}$, $s
= 683^{+69}_{-49}$, $V_{bw} = 12338$, $s_{b} = 685$, $V_{m} = 12386$,
$s_{m} = 635$ km s$^{-1}$, skewness 0.05, kurtosis -0.03. All further
conclusions are also referred to the sample with size $n = 72$.
   
\section{Results}
\label{sec:results}

In Fig.~2 we show $r_{max}$ results for the sub-samples $6
\leq \{i\} \leq 70$. 
For sample $\{70\}$ the correlation coefficients are 
$r_{max,70} = 0.358^{+0.055}_{-0.111}$, $r_{max,rand} = 0.154\pm 0.077$.
 Assuming that correlation coefficients $r_{max,70}$ and
$r_{max,rand}$ are almost normally distributed and are independent, the
significance of their difference 
$ \langle r_{max}\rangle - \langle r_{rand}\rangle $
can be estimated from Student's distribution. Consequently, the
probability of the null hypothesis, that the correlation coefficients
are drawn from one and the same population, is ${\cal P} < 1.0\cdot
10^{-13}$.

The highest $r_{max,47} = 0.637^{+0.061}_{-0.100}$
(where st.dev. are according to the MB method) occurs for sub-sample {47}, at
$d_{i} = 0.54$ Mpc. For B the st.dev. are (+0.069,-0.084).  
For the
randomised sub-sample {47} we have $r_{max,rand} = 0.209\pm0.104$.
In this case ${\cal P} < 1.0\cdot 10^{-17}$.

\begin{figure}
\includegraphics[width=60mm,angle=-90]{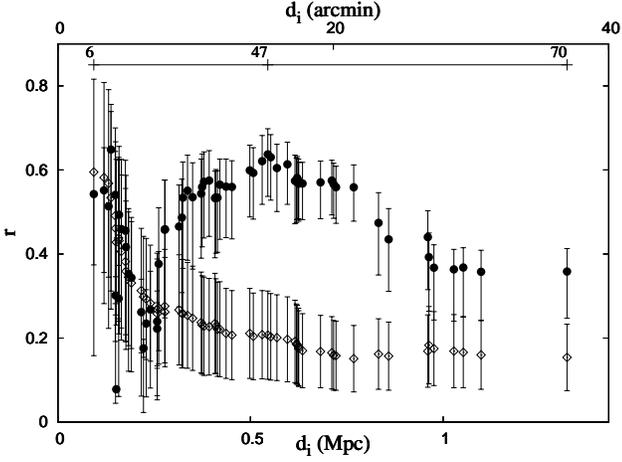}
 \caption{Observed $r_{max}$ (dots) and azimuthally randomised  
 $\langle r_{max,B} \rangle$ (diamonds) for subsamples $\{i\} = 6 \div 70$
 as a function of the distance from the cluster centre $d_{i}$. The st.dev.
 for $r_{max}$ are defined according to MB while for 
  $\langle r_{max,B} \rangle$ are according to B. These types of errors are 
  presented on all figures except Figs. 6-7.}
  \label{fig:rmax}
\end{figure}

A crucial test for rotation are the diagrams ($\varphi_{rmax},d_{i}$)
and ($\varphi_{r0},d_{i}$).  If the points 
are randomly distributed in the interval $[0\degr, 360\degr)$ the 
hypothesis for rotation must be rejected. The results for both 
$\varphi_{rmax}$ and $\varphi_{r0}$ are shown on Fig.~3, together 
with the corresponding results for azimuthally randomised 
sub-samples (denoted as stars). Note that $\varphi_{r0,rand} \equiv 
\varphi_{rmax,rand}$.  The st.dev. for the randomised sub-samples 
are $\approx [(360\degr)^{2}/12]^{1/2}$ which corresponds to the 
st.dev. of uniformly distributed positional angles. The differences 
between $\varphi_{r0}$ and $\varphi_{rmax}$ are $\approx 90\degr$.

\begin{figure}
\includegraphics[width=60mm,angle=-90]{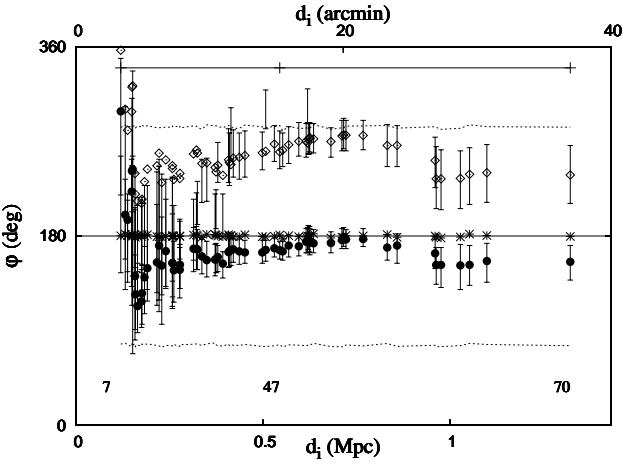}
 \caption{The positional angles $\varphi_{rmax}$ (dots) and $\varphi_{r0}$ 
 (diamonds). The stars show the results for azimuthally randomised sub-samples 
 with the errors (dotted lines).}
\label{fig:phi}
\end{figure}

Next, for sub-samples with $0.54<d_{i}\ Mpc<0.76$, for which $r_{max}$
significantly differs from the azimuthally randomised correlation
coefficients, $\langle\varphi_{rmax}\rangle = 173\degr \pm 4\degr$. 
We assume that this is the positional angle of the velocity gradient.
The positional angle of the minor axis is
$\langle\varphi_{r0}\rangle = 271\degr\pm5\degr$. The last value gives
the positional angle of the rotational axis if our model is valid. The
difference between both mean values is $98\degr$.

\begin{figure}
\includegraphics[width=60mm,angle=-90]{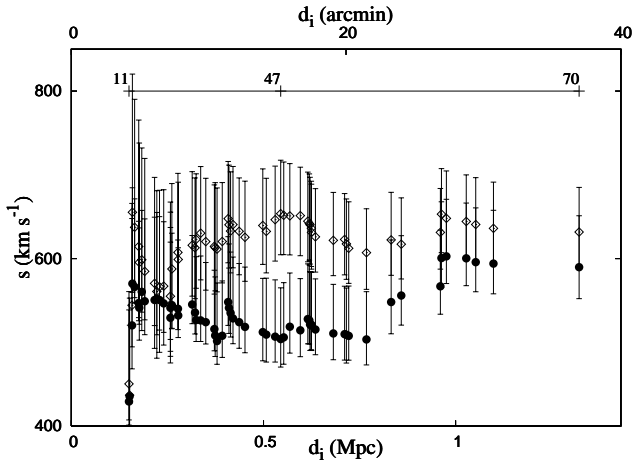}
 \caption{St.dev. of the regressions: $s_{r0}$ (dots) and $s_{rmax}$
  (diamonds).}
  \label{fig:reg}
\end{figure}

In Fig.~4 we show the st.dev. of the regressions $s_{rmax}$ and
$s_{r0}$ for positional angles corresponding to $r_{max}$ and
$r0$  -- there is significant distinction for those sub-samples where the
maximum correlation coefficient is significant.

\begin{figure}
\includegraphics[width=60mm,angle=-90]{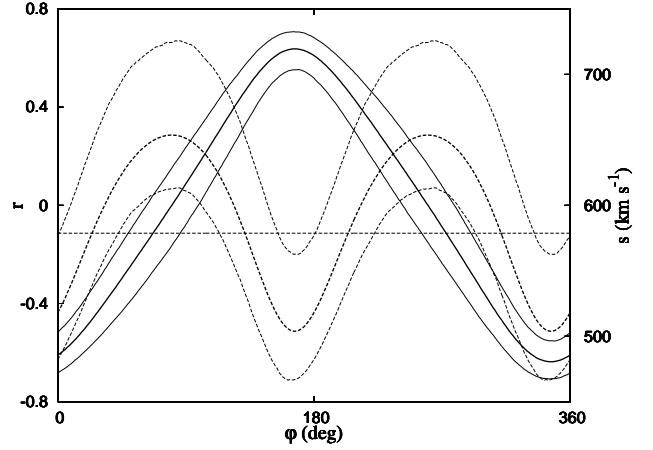} 
\caption{Variation of $r$ (solid line)  and $s$ (dashed) as a 
function of the positional angle $\varphi$ 
for sub-sample $\{i\} = 47$.} 
\label{fig:sine}
\end{figure}

The variations of $r$ and $s$ as a function of $\varphi$ for
sub-sample $\{i\} = 47$ are given on Fig.~5.  The curves are very
close to sine waves. For the maximum correlation coefficient
determined from azimuthally randomization of any sub-sample $\{i\}$,
the $s$ curve is also sine-like but with much smaller amplitude and
for various bootstrap resamples their phase angles are randomly
distributed on $[0\degr,360\degr)$.

For the sub-sample \{47\} we have $\varphi_{rmax} = 166\degr$ and
$\varphi_{r0} = 260\degr$.  And the virial mass estimate ${\cal M} =
(2.18 \pm 0.36) \cdot 10^{14} {\cal M}_{\sun}$ and ${\cal M}^{c} =
(1.29 \pm 0.21) \cdot 10^{14} {\cal M}_{\sun}$.
   
\begin{figure}
\includegraphics[width=60mm,angle=-90]{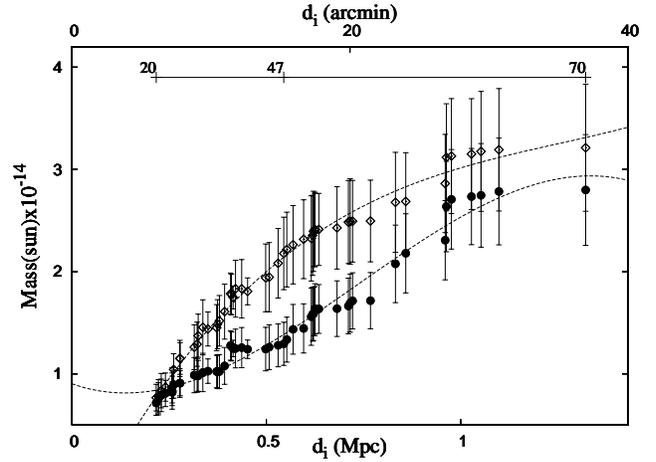}
 \caption{The virial masses: ${\cal M}^{c}$ (dots) and ${\cal M}$ (diamonds).
 The error bars refer to the jackknife estimator. Fits with third order 
 polynomials are also shown.}
 \label{fig:vir}
\end{figure}

We have computed the cumulative virial mass of A2107 for sub-samples
\{20\} to \{70\}, 
 following the prescriptions of Heisler, Tremaine \&
Bahcall (1985).  We have used the velocity dispersion $\sigma_{V}$ to
derive the usual virial mass ${\cal M}$ as well as the corrected
virial mass ${\cal M}^{c}$.  The results are given on Fig.~6.  The
error bars were calculated implementing the jackknife method. There is
a slight difference between ${\cal M}$ and ${\cal M}^{c}$. The
distinction begins at $\{i\} = 20$, at $d_{20} = 0.22$ Mpc. In order
to compare both estimates we fit the masses with third order
polynomials, as shown on Fig.~6. On Fig.~7 we show ${\cal M} - {\cal
  M}^{c}$ together with the difference between both polynomials. If
our finding that the cluster A2107 rotates is true, then Fig.~6 would
reflect a fictitious mass, due to rotation. Our estimates for the total
masses are ${\cal M} = (3.21\pm0.62)\cdot10^{14}{\cal M}_{\sun}$ and
${\cal M}^{c} = (2.80\pm0.54)\cdot10^{14}{\cal M}_{\sun}$.

\begin{figure}
\includegraphics[width=60mm,angle=-90]{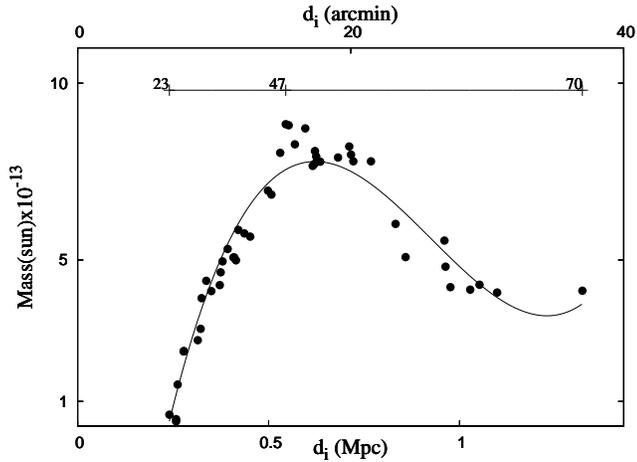}
 \caption{The difference between ${\cal M}$ and ${\cal M}^{c}$.}
  \label{fig:dmass}
\end{figure}

The parameter $\beta$ of the examined regression, which represents the
maximum velocity gradient expressed in km
s$^{-1}$ Mpc$^{-1}$ is shown on Fig.~8. Actually it is $\beta_{rmax}$
and also $\langle\beta_{rmax,MB}\rangle$ as functions of $d_{i}$.

Our method gives an opportunity to estimate an upper limit for the
rotational period. In Fig.~9 we show the upper limits for $P_{rmax}$
compared with the results from azimuthally randomised samples. The
estimate is simplistic -- we have not included any correction for the
inclination of the galaxy cluster.  Any inclination of our flat model
will inevitably decrease the period.

\begin{figure}
\includegraphics[width=60mm,angle=-90]{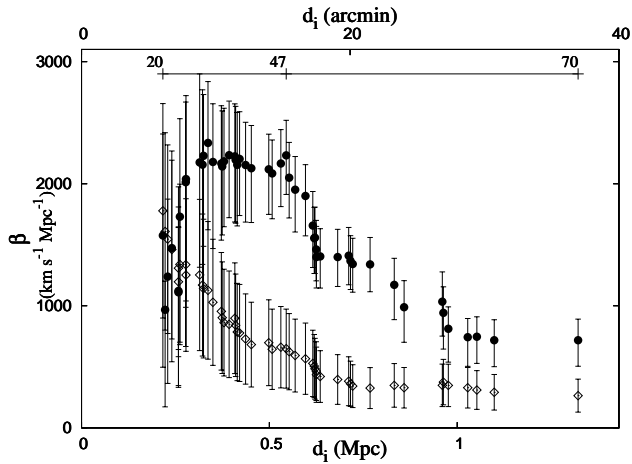}
 \caption{Parameters $\beta_{rmax}$ (dots) and
 $\langle \beta_{rmax,MB} \rangle$ (diamonds).}
  \label{fig:beta}
\end{figure}

\begin{figure}
\includegraphics[width=60mm,angle=-90]{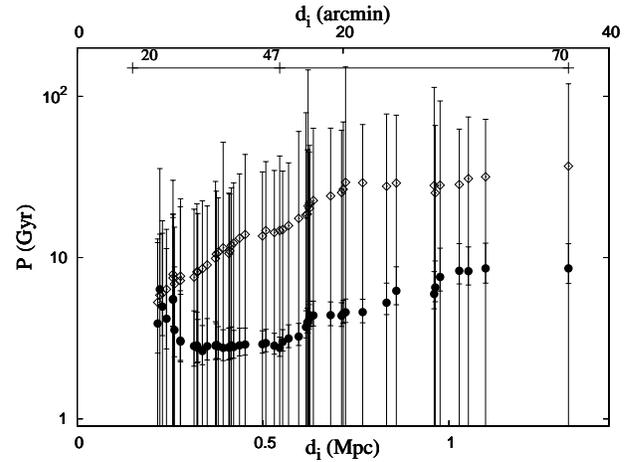}
 \caption{Upper limits of the period $P_{rmax}$ (dots) and
 $\langle P_{rmax,MB} \rangle$ (diamonds).}
  \label{fig:period}
\end{figure}

The analysis up to now was based on cumulative distributions.  It is
possible however to obtain an additional information by considering
the differential case. Let us denote the
differential sub-samples by $\{p,p+q-1\}$, where $p=1,2,\dots,70$ and
$q = 3, 4,\dots,70$. Repeating our procedure for the differential
sub-samples, which are in fact galaxies in different annuli, we find
that $r_{max}$ is very noisy for small $p$ and $q$ and at its maximum
$r_{max} = 0.814^{+0.040}_{-0.102}$ for $\{25,47\}$ (i.e. $p=25$ and
$q=23$ or for distances $0.26 \div 0.54$ Mpc from the cluster centre).  For
the same sub-sample the azimuthally randomised bootstrap resampling
gives $r_{max} = 0.271 \pm 0.133$. 
The corresponding probability is ${\cal P} < 1.0\cdot 10^{-11}$ and it 
is larger than
that for sub-sample $\{47\}$ since there are fewer degrees of freedom. 
 
The standard deviations of the estimates for the
$r_{max}$ and $r_{0}$ are $s_{rmax} = 446^{+118}_{-42}$ km s$^{-1}$ and
$s_{r0} = 767^{+92}_{-56}$ km s$^{-1}$ correspondingly. For the
differential sub-sample $\varphi_{rmax} = 167\degr$ and $\varphi_{r0}
= 261\degr$.  For these 23 galaxies the derived period is $P =
(2.43^{+0.30}_{-0.21}) \cdot 10^{9}$ years. The corresponding
differential virial masses are ${\cal M} =(3.51\pm0.60) \cdot
10^{14}{\cal M}_{\sun}$ and ${\cal M}^{c} = (1.19\pm0.20) \cdot
10^{14}{\cal M}_{\sun}$.
 
The results from the differential treatment are in agreement and
support the cumulative ones.

 \section{Discussion and conclusions}
 \label{sec:final}

A2107 is a nearby cluster of richness class 1, classified as BMI
(Abell 1958, Abell, Corwin \& Olowin 1989). The Abell's count of
galaxies is $N_{A} = 51$, the cluster radius is 45 arcmin (22.4 arcmin from
Struble \& Rood 1987, $H_{0} = 50$) and the RS
classification type is cD (Struble \& Rood 1987). The cluster is
isolated and is also identified (Kalinkov, Valtchanov \& Kuneva 1998b)
 as Zwicky cluster ZC~7573 (Zwicky \&
Herzog, 1963) as distance group {\em Near}, type {\em mc}, population
of galaxies 293 and equivalent radius $r_{eZ} = 69.4$ arcmin. 
At $26.5\arcmin$  and positional 
angle of $204\degr$ is located another Zwicky cluster ZC~7578 (distance 
group {\em  VD}, population 110, type {\em mc}, $r_{eZ} = 7.8$ arcmin).

There is no detailed description of A2107. Girardi et al. (1997)
comment on A2107 -- ``Remarkably regular cluster''.
 
Presumably rotation of clusters of galaxies is difficult to be
established with certainty. 
Materne \& Hopp (1983) have shown that it is extremely hard (if
generally possible) to distinguish the case of a single cluster in
rotation from the case of two overlapping clusters, which are merging
or departing from each other. That is why OH, investigating the same
cluster A2107, have examined both possibilities. They employed the
$\delta$-test of Dressler \& Shectman (1988) and they have found that
there is significant sub-structuring in A2107.  Supposing that
$\sum_{l=1}^{n}\delta_{l} \approx n$, when there is no
spatial-velocity correlation, OH performed 1000 Monte Carlo (MC)
simulations and found $\langle\sum_{l}\delta_{l}\rangle_{MC} = 111$
(see also Oegerle \& Hill 2001). Because their sample contains $n=68$
cluster members, they concluded that A2107 contains
substructures. All $\sum\delta_{l,MC}$ are smaller than 111 and hence
the probability for existence of substructures in A2107 is $>
0.999$. We have carried out exactly the same experiment but with
$10^{4}$ Monte Carlo simulations and found that
$\sum_{l}\delta_{l,MC}$ are almost normally distributed with mean
$\langle\sum_{l}\delta_{l}\rangle_{MC} = 70.00\pm10.82$ against $\sum
\delta_{l,obs} = 102.3$.  Our result support the conclusion of OH that
A2107 indeed contains substructures with probability $> 0.994$.

Concerning the velocity distribution, relying on the large peculiar
velocity of the central cD galaxy (UGC~9958, $V_{pec} = 270$ km
s$^{-1}$; see also Oegerle \& Hill, 2001), OH decomposed the
velocity histogram into two sub-clusters. They have chosen one
decomposition amongst infinite number of possibilities. Indeed, there
are no reasonable arguments that constrain us to assume two
sub-cluster centres or that the velocity histogram is superposition of
two (or may be more?) distributions.

We have verified the hypothesis for sub-clustering -- one against two
centres.  We have used another powerful test especially for two
sub-structures (Lee 1979), introduced in the cluster sub-structure
studies by Fitchett (1988) and successfully applied by Fitchett \&
Webster (1987), Rhee, van Haarlem \& Katgert (1991b), Pinkney et al. 
(1996). Our result, on the base of 10 000 simulations of
azimutally randomised samples, leads to Lee function $L_{az} =
1.42\pm0.21$ while $L_{obs} = 2.11^{+0.31}_{-0.41}$. The alternative
hypothesis, that the cluster has two sub-structures, is accepted at
significance of at least 95\%. The corresponding positional angle of
the two sub-cluster centres is $145\degr$.

But the subclustering does not rule out the hypothesis of rotation. It
is plausible that the cause of sub-clustering in the Dressler-Shectman
diagram (Fig.~5 of OH) is just due to the rotation, since the most
prominent clumpiness is along the maximum gradient in the velocity
field of A2107, indicated by our $\varphi_{rmax}$. Indeed, from galaxy
studies we know that the velocity field of disk-like rotating galaxies
without any massive halo has two extremes, $V_{max}$ and $V_{min}$.
This would cause apparent sub-clustering. But if there are two
overlapping clusters, then it will be unlikely they will generate a
mimicry of rotation -- first of all, the velocity histogram will have
two peaks or quite broad velocity distribution with rather
unrealistically high velocity dispersion. None of this is observed in
A2107. Secondly, we have
investigated the behaviour of correlation coefficients,
positional angles etc. From our point of view, if there are two
sub-clusters that are located at positional angle close to
$\varphi_{rmax}$, this would be regarded as evidence for
rotation. Nevertheless some authors disregard this possibility as the
cause of the velocity gradient (den Hartog \& Katgert 1996).

Our strongest evidence for rotation is the consistent measure of the
positional angle of the velocity gradient $\varphi_{rmax}$ for
consecutive cumulative sub-samples. One way of validating the inferred
positional angle is to compare it with the positional angle of the
elongation of the cluster $\varphi_{el}$.  For our flat disk-like
model $\varphi_{el} = \varphi_{rmax} = \varphi_{a}$.  Debating the
Binggeli's effect (Binggeli 1982) many authors have measured
$\varphi_{el}$, as well as ellipticities, of various samples of Abell
clusters (e.g. Struble \& Peebles 1985, Struble \& Rood 1987, Lambas
et al.  1990). While Struble \& Peebles (1985) have found
$\varphi_{el} = 137\degr$ for A2107, Rhee \& Katgert (1987) quote $160
\degr$ and $93\degr$ for the 50 and 100 brightest galaxies
respectively, within a radius of 1 Mpc. According to Binggeli (1982)
$\varphi_{el} = 62\degr$. In a detailed investigation Rhee, van Haarlem
\& Katgert (1991a, 1992) apply three different methods and determine
for the same cluster, using about 300 galaxies out to 0.75 Mpc, that
$\varphi_{el} = 75, 68$ and $90\degr$ with ellipticities 0.13, 0.00
and 1.00 respectively. But according to their Fig.~5 (Rhee, van
Haarlem \& Katgert 1991a) $\varphi_{el} \approx 143\degr$ for the
first 20 brightest galaxies while for the first 70 is about
$3\degr$. 

Apparently the estimates of $\varphi_{el}$ for A2107 are not
very reliable. Perhaps the reason is objective -- it seems that
$\varphi_{el}$ depends on the brightness of the galaxies and the
distance to the cluster centre. There are examples of drastic
variations of $\varphi_{el}$ in some clusters in the most elaborate
paper on this subject (Burgett et al. 2004).

As stated above there are two possibilities to explain the velocity gradient 
in a galaxy cluster - two-body model or rotation. An anonymous referee however
turned our attention on a third possibility: according to
simulations of galactic tides in dwarf spheroidal galaxies Piatek \& Pryor 
(1995) have shown that tides produce large ordered motions which induce apparent
rotation. Biviano et al. (1996) assume that the velocity gradient in the 
central part of Coma cluster maybe due to tidal effects caused by falling 
groups of galaxies and not by rotation. 
But in the case of A2107 it is not quite the same,
because we did not find any groups of galaxies in this regular
cluster.

The parameters $\beta$ and $\varphi_{rmax}$ derived with our method
may be compared with those of other authors.
The positional angle of the rotational axis of OH, namely $70\degr$, is close
to our determination $271\degr - 180\degr = 91\degr$.
From $\partial V/\partial X, \partial V/\partial Y$
for A2107 in Table 3 of den Hartog \& Katgert (1996) we infer
$\varphi_{rmax} \approx 330\degr$. 
The last value is close to our
estimate of $\varphi_{rmax} = 173\degr$ with
$180\degr$ difference depending on the convention used.

The parameter $\beta$
after den Hartog \& Katgert (1996) is 842 km s$^{-1}$ Mpc$^{-1}$ (for
$H_{0} = 50$ km s$^{-1}$ Mpc$^{-1}$).  Our value $\beta$ for the 
entire cluster A2107 is 718 while for the subsamples $\{47\}$ and $\{25,47\}$
the values are 2232 and 2527 km s$^{-1}$ Mpc$^{-1}$ correspondingly. 
OH give $\approx 1800$ km s$^{-1}$ Mpc$^{-1}$ for the
whole cluster, which is close to our value for the sub-samples but not
for sample $\{70\}$.

Our estimate of the rotational period is $P =
\left(2.4^{+0.3}_{-0.2}\right) \cdot 10^{9}$ years, while the
azimuthally randomised bootstrap generations give $P=
(1.3\pm2.3)\cdot 10^{10}$ years. We have to note that our method
of searching the maximum correlation in random samples leads to
artificially decreasing the period of rotation. The period quoted by OH
is $7\cdot 10^{9}$ years. These periods are not corrected for the
inclination of the cluster. 
Our estimate of $P$ could be compared with the corresponding value of Gregory 
\& Tifft (1976) for the outer regions in Coma cluster (A1656), which is
about $2 \cdot 10^{11}$ years.
We presume the real period is $< 2\cdot
10^{9}$ years and therefore A2107 has made already few rotations
during the Hubble time.

There are several virial mass estimates for A2107. All estimates are
not substantially different because they are based on one and the same
radial velocity data. According to OH the total virial mass of A2107
is ${\cal M} \approx 4\cdot 10^{14}{\cal M}_{\sun}$ and in the frame
of their two-body model, the masses of both sub-structures are $\sim
1.3:1$. Girardi et al. (1998) give a virial mass ${\cal M} = \left(
3.02^{+0.94}_{-0.89}\right)\cdot 10^{14}{\cal M}_{\sun}$ and a
corrected ${\cal M}^{c} = \left( 2.62^{+0.83}_{-0.77}\right)\cdot
10^{14}{\cal M}_{\sun}$ (according to their Eqn.~8). Girardi et
al. (2002) use the corrected mass only. But the correction of Girardi
et al. (1998, 2002) has quite a different nature -- it is due not to
rotation but to a modification of the virial theorem in order to
compensate the incompleteness of the sample out to the virial radius.

Our estimated total mass of A2107 is ${\cal M} = (3.21\pm0.62)\cdot
10^{14}{\cal M}_{\sun}$ (without any correction). Most importantly,
reducing the effect of rotation we infer virial mass ${\cal M}^{c} =
(2.80\pm0.54)\cdot 10^{14}{\cal M}_{\sun}$. 
The simplest way to get this
virial mass is to estimate the velocity dispersion from the linear
regression at positional angle $\varphi_{r0}$.

\section*{Acknowledgements}

We are very grateful to an anonymous referee for constructive remarks, 
useful suggestions and helpful comments which have significantly improved 
the paper.

This research has made use of the NASA/IPAC Extragalactic Database
(NED), which is operated by the Jet Propulsion Laboratory, California
Institute of Technology, under contract with the National Astronautics
and Space Administration. This research has made use of Aladin
(Bonnarel et al. 2000) and of the MAPS Catalog of POSSI which are
supported by the National Aeronautics and Space Administration and the
University of Minnesota. The APS databases can be accessed at
http://aps.mn.edu/. We are very thankful to T. Beers for providing us
with ROSTAT package.

\bsp

\label{lastpage}

\end{document}